\documentclass[twocolumn,prc,showpacs,amsmath,amssymb,superscriptaddress,floatfix]{revtex4-1}

\usepackage{graphicx}
\usepackage{dcolumn}
\usepackage{bm}

\begin{document}

\title{
Novel shape evolution in exotic Ni isotopes and configuration-dependent 
shell structure  
}
\author{Yusuke Tsunoda}
\affiliation{Department of Physics, University of Tokyo, Hongo, 
Bunkyo-ku, Tokyo 113-0033, Japan}
\author{Takaharu Otsuka}
\affiliation{Department of Physics, University of Tokyo, Hongo,
Bunkyo-ku, Tokyo 113-0033, Japan}
\affiliation{Center for Nuclear Study, University of Tokyo, Hongo,
Bunkyo-ku, Tokyo 113-0033, Japan}
\affiliation{National Superconducting Cyclotron Laboratory, 
Michigan State University, East Lansing MI 48824, USA}
\author{Noritaka Shimizu}
\affiliation{Center for Nuclear Study, University of Tokyo, Hongo,
Bunkyo-ku, Tokyo 113-0033, Japan}
\author{Michio Honma}
\affiliation{Center for Mathematical Sciences, University of Aizu, 
Ikki-machi, Aizu-Wakamatsu, Fukushima 965-8580, Japan}
\author{Yutaka Utsuno}
\affiliation{Advanced Science Research Center, Japan Atomic Energy
Agency, Tokai, Ibaraki 319-1195, Japan}
\date{\today}

\begin{abstract}
\noindent
The shapes of neutron-rich exotic Ni isotopes are studied.   
Large-scale shell model calculations are performed by advanced Monte Carlo Shell Model 
(MCSM) for the $pf$-$g_{9/2}$-$d_{5/2}$ model space.  Experimental 
energy levels are reproduced well by a single fixed Hamiltonian.  Intrinsic shapes are 
analyzed for MCSM eigenstates.  Intriguing interplays among spherical, oblate, prolate and
$\gamma$-unstable shapes are seen, including shape fluctuations, $E$(5)-like situation,  
the magicity of doubly-magic $^{56,68,78}$Ni, and the coexistence of spherical and 
strongly deformed shapes.   Regarding the last point,    
strong deformation and change of shell structure can take place simultaneously, 
being driven by the combination of the tensor force and changes of major configurations
within the same nucleus.       

\end{abstract}


\maketitle

Atomic nuclei exhibit simple and robust regularities in their structure comprised of 
$Z$ protons and $N$ neutrons.   A very early example is the (spherical) magic numbers 
conceived by Mayer and Jensen \cite{MJ}.
These magic numbers dominate low-energy dynamics of stable nuclei and 
their neighbors on the Segr\'e chart. 
Another basic feature is nuclear shape, which has been one of the central issues
of nuclear physics, since Rainwater \cite{R} and Bohr and Mottelson \cite{BM}.  
The shape varies as $Z$ or $N$ changes in such a way that  
it tends to be spherical near magic numbers, while it becomes more deformed 
towards the middle of the shell.   
Recent theoretical and experimental studies on exotic nuclei with unbalanced $Z$ and $N$ 
cast challenges to these pictures.
Even magic numbers are not an exception: the changes of the shell structure 
due to nuclear forces, referred to as {\it shell evolution} \cite{nobel}, have been seen,  
including disappearance of traditional magic numbers and appearance of new ones.  
A recent example is the discovery of the $N$=34 magic number \cite{54Ca}, 
after its prediction a decade ago \cite{magic}, 
while many other cases have been discussed 
\cite{nobel,review,tensor,vmu}.  

It is, thus, of much interest to explore shapes 
of exotic nuclei and to look for relations to the shell evolution.
In this Rapid Communication, we report results of state-of-the-art large-scale shell model calculations
for a wide range of Ni isotopes, focusing on these points. 
While the ground state turns out to be basically spherical, 
a strongly prolate deformed band appears at low excitation energy in some nuclei, 
similar to shape coexistence, known in other nuclei over decades, {\it e.g.}  
\cite{morinaga,coexist_pt186,coexist}.
We shall present that the shell structure, for instance, the spin-orbit splitting, can be 
varied significantly between such spherical and deformed states by a combined effect
of different major configurations and the nuclear forces, particularly the proton-neutron tensor 
force.  This phenomenon occurs within the same nucleus, and thereby is not described as the
shell evolution in the conventional sense.  However, in order to discuss the basic underlying physics in a  
unified way, this phenomenon will be called {\it Type II shell evolution}, while 
the shell evolution by the change of $N$ or $Z$ will be referred to as {\it Type I}.     
We shall discuss other interesting features, {\it e.g.}, varying appearance of magicity in 
$^{56,68,78}$Ni, shape fluctuations including $\gamma$ instability, and the $E$(5)-like case \cite{E5}.
 
We discuss, in this Rapid Communication, the structure of Ni isotopes of even $N$=28$-$50, 
utilizing results of the advanced Monte Carlo Shell Model (MCSM) 
calculation \cite{mcsm_review,mcsm_extra,mcsm_old} run on the $K$ computer for $\sim$2$\times 10^{10}$ 
core seconds in total.   The model space consists 
of the full $pf$ shell, $0g_{9/2}$ and $1d_{5/2}$ orbits for both protons and neutrons.
There is no truncation within this space, as an advantage of MCSM.  
The Hamiltonian is based on the A3DA Hamiltonian with minor revisions \cite{mcsm_review,a3da}.  
The spurious center-of-mass motion is removed by the Lawson method \cite{lawson}.

\begin{figure}[tb]
 \begin{center}
 \includegraphics[width=7cm,clip]{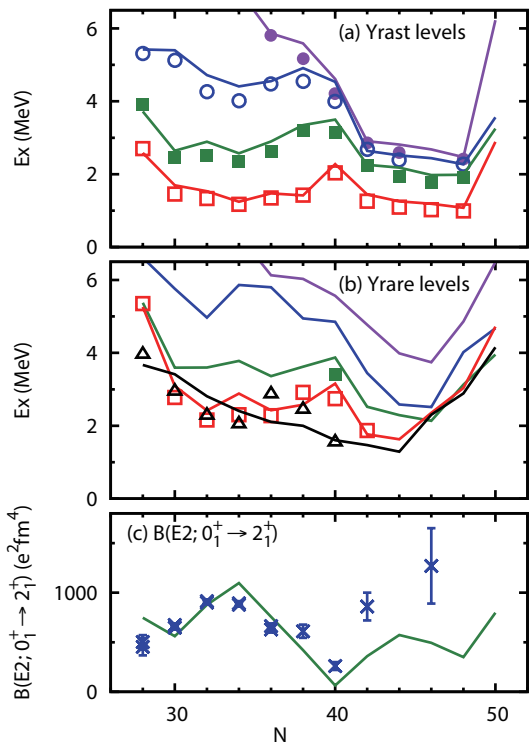}
 \caption{
(Color online) Energy levels for (a) yrast and (b) yrare states of Ni isotopes with even $N$.  
Symbols are experimental data for $J^{\pi}$=0$^+$ (black triangle), 2$^+$ (open red square), 
4$^+$ (green filled square), 6$^+$ (open blue circle) and 8$^+$ (filled purple circle) 
\cite{ensdf,ni68_recchia,ni68_suchyta}.   
Lines are present MCSM calculations with the same color code.
(c) $B(E2;0^+_1 \rightarrow 2^+_1)$ values from experiment \cite{be2_exp} and by the present
calculation. 
}
 \label{fig1}
 \end{center}
\end{figure}

Figure \ref{fig1} shows yrast and yrare levels by the present calculation compared to experiment 
\cite{ensdf,ni68_recchia,ni68_suchyta}.
Systematic behaviors are visible in experimental yrast levels as well as $J^{\pi}$=0$^+_2$  
and 2$^+_2$ yrare levels, with a remarkable agreement to the theoretical trends.    
Such good agreement has been obtained with a single fixed Hamiltonian, and suggests 
that the structure of Ni isotopes can be studied with it.
The $B(E2;0^+_1 \rightarrow 2^+_1)$ values 
with neutron and proton effective charges, 0.5 and 1.5, respectively, 
are shown in Fig. \ref{fig1} compared to 
experiment \cite{be2_exp} with certain discrepancies for heavier isotopes, 
where uncertainties are larger and (p,p') data is converted ($N$=46) \cite{ni70_perru,aoi}.  
A more systematic comparison with precise data is desired.
Relevant shell model calculations have been reported \cite{lenzi,kaneko}.  In particular, those of 
\cite{lenzi} are a remarkable achievement of the large-scale conventional shell-model approach, with good    
agreement to experiment.  Many experimental data are yet to be obtained.  
For instance, the 0$^+_2$ level of $^{68}$Ni has only recently been corrected  
\cite{ni68_recchia,ni68_suchyta}.
The primary objective of this Rapid Communication is to predict novel systematic change of band structures  
in $^{68-78}$Ni isotopes and to present the under-lying robust mechanisms for them. 

\begin{figure}[tb]
 \begin{center}
 \includegraphics[width=8.5cm,clip]{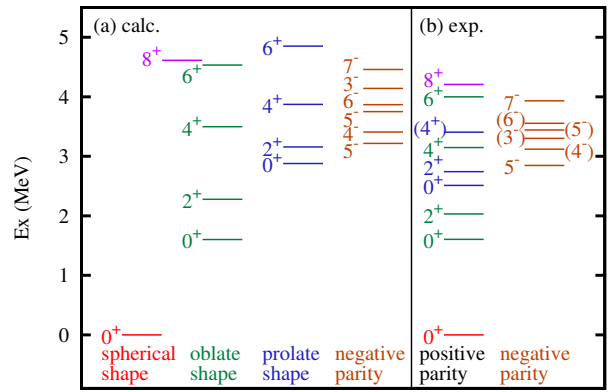}
 \caption{
(Color online) Energy levels of $^{68}$Ni by present calculation (left panel) and 
by experiment (right panel) \cite{ensdf,ni68_recchia,ni68_suchyta}.
}
 \label{fig2}
 \end{center}
\end{figure}

We show in Fig. \ref{fig2} a more detailed level scheme for $^{68}$Ni, including negative-parity 
states.   This nucleus has attracted much attention \cite{ni68_recchia, ni68_suchyta,kaneko,lenzi,
ni68_girod, ni68_ishi, ni68_sorlin,ni68_langanke, ni68_pauwels, ni68_dijon, ni68_chiara, ni68_broda_2012}
from both theoretical and experimental sides.
The positive-parity levels are classified according to their shape categories: spherical, 
oblate and prolate.  We shall come to this point later.
The correspondence between theoretical and experimental levels can be made with rather 
good agreement, including levels of higher spins.   

\begin{figure*}[tb]
 \begin{center}
 \includegraphics[width=16cm]{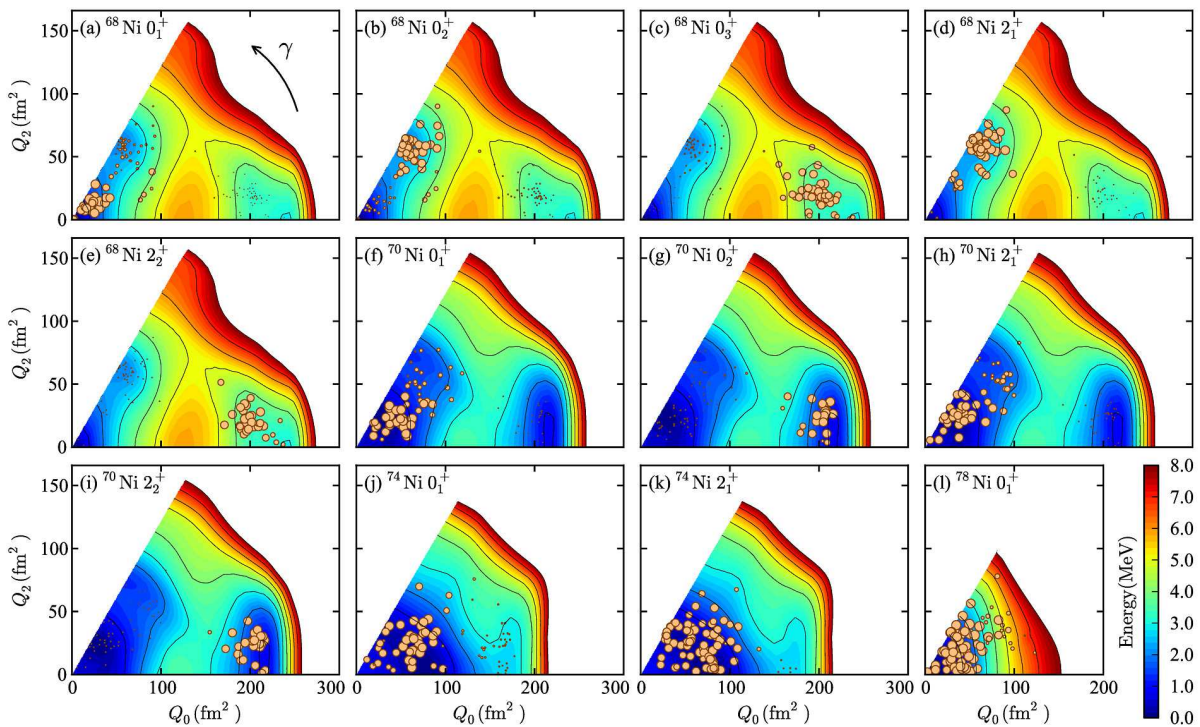}
 \caption{
(Color online) Potential energy surfaces (PES) of Ni isotopes, coordinated 
by usual $Q_0$ and $Q_2$ (or $\gamma$).
The energy relative to the minimum is shown by contour plots.
Circles on the PES represent shapes of MCSM basis vectors (see the text).
}
 \label{fig3}
 \end{center}
\end{figure*}

Figure \ref{fig3} depicts, for selected states of $^{68,70,74,78}$Ni isotopes, 
potential energy surface (PES) for the present Hamiltonian
obtained by the Constraint Hartree-Fock (CHF) method with the usual constraints on the quadrupole moments
$Q_0$ and $Q_2$.
We can see many features: for instance, for $^{68}$Ni, there is a spherical minimum 
stretched towards modest oblate region, as well as a prolate local minimum.

The MCSM wave function is expressed by a superposition of Slater determinants with 
the angular-momentum and parity projector $P[J^{\pi}]$,
\begin{equation}
\Psi = \sum_{i} c_i P[J^{\pi}] \, \Phi_i \, .
\label{eq1}
\end{equation}
Here, $c_i$ denotes an amplitude, 
$\Phi_i$ stands for the Slater determinant consisting of one-nucleon wave functions 
$\phi^{(i)}_1, \phi_2^{(i)}, ...,\phi^{(i)}_n$ with
\begin{equation}
\phi^{(i)}_k = \sum_{l} D^{(i)}_{k,l} \, u_l ,
\label{eq2}
\end{equation}  
where $u_l$ is the $l$-th single-particle state in the original model space in $m$-scheme,  
and $D$ implies an amplitude determined by the MCSM process.  $\Phi_i$ is 
the product of the proton and neutron sectors, with $n$ being  
the number of valence protons or neutrons.  

For each $\Phi_i$, we take the following procedure.  We calculate its quadrupole moment matrix, 
and diagonalize it.   Three axes are obtained with $Q_0$ and $Q_2$ values.
We then place a circle on the PES at the point corresponding to these $Q_0$ and $Q_2$ values.  
The size ({\it i.e.} area) of the circle is set to be proportional to the overlap probability 
between $\Psi$ and the normalized $P[J^{\pi}] \, \Phi_i$. 
Thus, the location of the circle implies the intrinsic shape of $\Phi_i$, 
and its size the importance of it in the eigenstate, $\Psi$.   
Note that the states $P[J^{\pi}] \, \Phi_i$ ($i$=1, 2, ...) 
are not orthogonal to each other, in general, but the distribution pattern of the circles provides 
a unique and clear message on the intrinsic shape of the shell-model eigenstate, as we shall see. 

\begin{figure}[tb]
 \begin{center}
 \includegraphics[width=6.5cm,clip]{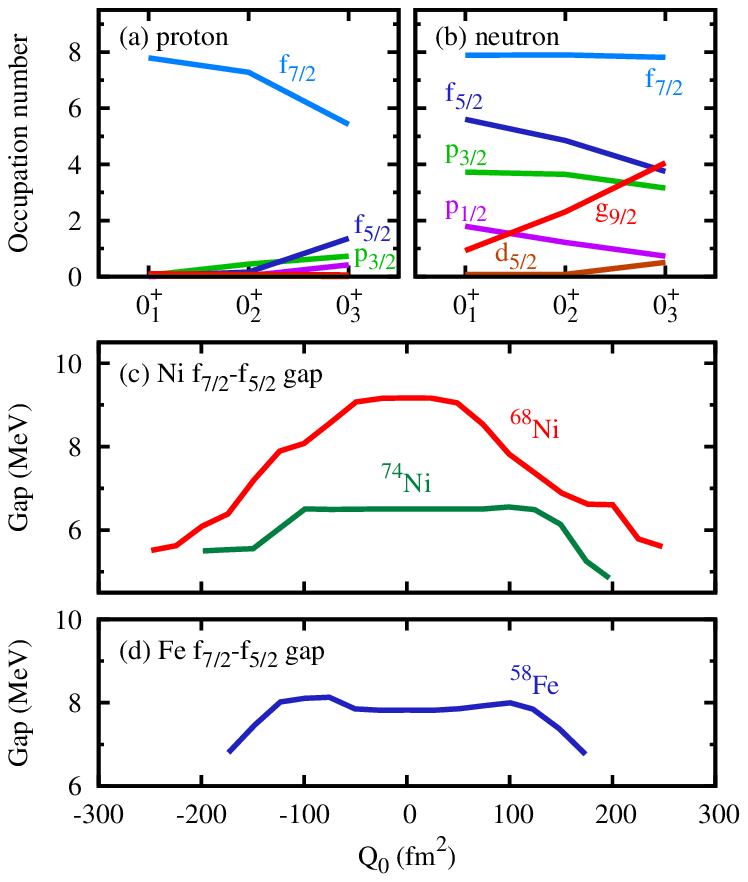}
 \caption{
(Color online) (a,b) Occupation numbers for the 0$^+$ states of $^{68}$Ni.
Gap between $f_{7/2}$ and $f_{5/2}$ as a function of $Q_0$ of Fig.~\ref{fig3}
for (c) $^{68,74}$Ni and (d) $^{58}$Fe.  
}
 \label{fig4}
 \end{center}
\end{figure}

Figure \ref{fig3}(a) shows such circles for the ground state of $^{68}$Ni.  We see many 
large circles near the spherical point, $Q_0$=$Q_2$=0.   
In general, there can be many points close to one another partly because 
each circle represents a Slater determinant and a two-body 
interaction, particularly its pairing components, mixes different Slater determinants.
Those Slater determinants should have similar shapes so that the mixing between 
them can occur. 
We also see notable spreading of the distribution of circles from the spherical point.    
This implies the extent of the shape fluctuation.  
The 0$^+_2$ state in Fig. \ref{fig3}(b) shows similar spreading but the locations are 
shifted to the moderately oblate region ($\beta_2 \sim -0.2$).
Although there is no clear potential barrier between the spherical and oblate regions of the PES, 
the antisymmetrization pushes the 0$^+_2$ state away from the 0$^+_1$ state.
Figure \ref{fig3}(c) exhibits many circles in a profound prolate minimum with 
$Q_0 \sim $200 fm$^2$ ($\beta_2 \sim$ 0.4).   
We emphasize that we can analyze, in this way, the intrinsic shape even for 0$^+$ states without
referring to E2 properties.   

Figures \ref{fig3}(d,e) show the same plots for the 2$^+_{1,2}$ states.  
The 2$^+_{1}$  state exhibits a pattern almost identical to that of 
the 0$^+_{2}$ state, which suggests the formation of the modestly-oblate band.   
Such striking similarity is found also between the 0$^+_{3}$ and 2$^+_{2}$ states with a 
strong-prolate-band assignment.   
The band structure can be further verified by E2 matrix elements,   
and is presented in Fig.~\ref{fig2} including 4$^+$ and 6$^+$ members.
We note that the 0$^+_{3}$ and 2$^+_{2}$ states of $^{68}$Ni were reported to be 
strongly deformed with $\beta_2 \sim 0.4$ in shell-model calculations in \cite{ni68_dijon}.

Figures \ref{fig4}(a,b) show occupation numbers of proton and neutron orbits, respectively, 
for the 0$^+_{1,2,3}$ states of $^{68}$Ni.  One sees drastic changes between the 
0$^+_{1}$ and 0$^+_{3}$ states for proton $f_{7/2}$ and neutron $g_{9/2}$, while 
some other orbits show also sizable changes.   Such changes are due to 
particle-hole excitations: mainly proton excitations from $f_{7/2}$ to $f_{5/2}$ and $p_{3/2,1/2}$, 
and neutron excitations from $f_{5/2}$ and $p_{1/2}$ to $g_{9/2}$.   Once such 
excitations occur, the state can be deformed towards an ellipsoidal shape and 
large deformation energy is gained predominantly from the proton-neutron 
quadrupole interaction.  
The configuration structure of the 0$^+_3$ state seems to be beyond    
the applicability of truncated shell model calculations \cite{jun45,jj44b}.

We next discuss effective single-particle energy (ESPE), obtained from the monopole 
component, $H_m$, of the Hamiltonian    
(see for instance \cite{nobel} for more details).   
$H_m$ is written in terms of the number operator, $n_j$, of each orbit $j$ (proton or
neutron is omitted).    
The ESPE is calculated usually for configurations that are being filled, but we evaluate it for 
mixed configurations by a functional derivative,
$\epsilon_j$= 
$\langle \frac{\partial H_m}{\partial n_j} \rangle$
with the expectation values of 
$n_j$'s for eigenstates being considered  
\footnote{
The contribution of identical particles in the same orbit becomes slightly 
different from the one by the filling scheme, but this difference is negligible 
in the present case.}.  
These $\epsilon_j$'s are still {\it spherical} ESPEs, but are obtained with $\langle n_j \rangle$
of deformed states.  
From the viewpoint of the Nilsson model, $\epsilon_j$'s correspond to 
Nilsson levels at the spherical limit, but the difference from the Nilsson model is that the $\epsilon_j$'s  
vary as the deformation changes, 
due to the orbit-dependence of the monopole component of nuclear forces. 
For protons, the ESPE of $f_{7/2}$ is increased by $\sim$2 MeV in going  from 0$^+_{1}$ to 
0$^+_{3}$ states, while ESPE of $f_{5/2}$ comes down by $\sim$1 MeV.  
Let us look into how these changes occur, based on the mechanism presented in 
\cite{tensor,nobel} :  
Because $g_{9/2}$ and $f_{7/2}$ are of $j_> (=l+1/2)$ type and $f_{5/2}$ is of $j_< (=l-1/2)$ type, 
the $g_{9/2}$-$f_{7/2}$ ($g_{9/2}$-$f_{5/2}$) monopole interaction from the tensor force is 
repulsive (attractive).   More neutrons in $g_{9/2}$ in the 0$^+_3$ state 
result in the raising of the proton $f_{7/2}$ and the lowering of the proton $f_{5/2}$.  Similarly, 
neutron holes in $f_{5/2}$ lead to the weakening of the attractive (repulsive) effect on the proton 
 $f_{7/2}$ ($f_{5/2}$).
All these effects reduce coherently the proton $f_{7/2}$-$f_{5/2}$ gap, ({\it i.e.}, the difference of the ESPEs of 
these orbits), making it $\sim$3 MeV 
narrower in the 0$^+_3$ state, including other minor effects.  

If a relevant shell gap becomes smaller, more particle-hole excitations occur 
over this gap, leading to stronger deformation with more energy gain  
as mentioned above.  
A stronger deformation enhances particular configurations, for instance, 
more neutrons in $g_{9/2}$, which reduce the proton $f_{7/2}$-$f_{5/2}$ gap further.
Thus, the change of the shell gap and strong deformation are interconnected 
in a self-consistent way.   
Figure ~\ref{fig4}(c) demonstrates this mechanism with an example of the 
proton $f_{7/2}$-$f_{5/2}$ gap 
obtained for the CHF wave function along the $\gamma=0 ^\circ$ and $60 ^\circ$ lines 
in Fig.~\ref{fig3}, as a function of $Q_0$.   
For $^{68}$Ni, as $Q_0$ departs from zero, the gap remains large, 
producing a high barrier between spherical and deformed shapes.
It, however, starts to come down beyond $|Q_0|  \sim $50 fm$^2$, and 
is lowered by 3 MeV for $|Q_0| \sim $200 fm$^2$.  Although this is a 
consequence of strong deformation, it also enhances the deformation.
Thus, the formation of a strongly prolate deformed band occurs by way of a 
non-linear mechanism, which stabilizes this band from spherical shape, 
by keeping the barrier high.  

We now compare the phenomenon discussed above to the shell evolution 
which occurs in many cases as a function of $Z$ or $N$  \cite{nobel,tensor,vmu}.  
Figure~\ref{fig4}(c) shows the proton $f_{7/2}$-$f_{5/2}$ gap of $^{74}$Ni also.  
It starts with $\sim$6 MeV at $Q_0$=0, much lower than the corresponding 
gap of $^{68}$Ni.
This reduction of the gap is nothing but a consequence of the shell evolution
with many neutrons occupying $g_{9/2}$.
We shall call it Type I shell evolution, hereafter.  Correspondingly,  
phenomena like the reduction of the proton $f_{7/2}$-$f_{5/2}$ 
gap of $^{68}$Ni at large deformation discussed above 
can then be called Type II shell evolution.  Type I and II shell 
evolutions occur due to the occupation changes of single-particle orbits,
while the changes are due to varying $Z$ and/or $N$ in Type I, but 
are due to particle-hole excitations within the same nucleus in Type II.
The tensor force plays important roles in both cases as it is the primary origin 
of the change of the spin-orbit splitting.

Figure~\ref{fig4}(c) shows that the $f_{7/2}$-$f_{5/2}$ gap of $^{74}$Ni 
stays constant up to  $|Q_0| \sim $120 fm$^2$, contrary to rapid change in 
$^{68}$Ni.  
Namely, Type II shell evolution is suppressed in $^{74}$Ni, as
$g_{9/2}$ is occupied by many neutrons causing Type I shell evolution.
Figure~\ref{fig3}(j) shows the PES of $^{74}$Ni, where the profound local minimum
of strong prolate deformation cannot be found consistently with the missing effect of
Type II shell evolution.
Such an interplay between Type I and II is of interest.  

The same gap is shown for $^{58}$Fe in Fig.~\ref{fig4}(d) as a typical 
standard case.  The pattern is quite flat similar to $^{74}$Ni.

Type II shell evolution can thus be introduced as a mechanism not by the change 
of $Z$ or $N$, but by the change of major configurations.
Type II shell evolution should enhance the appearance of the shape coexistence by stabilizing an isolated 
deformed local minimum.  Further studies in various situations are of extreme interest.       
We stress that Type II shell evolution can occur in other nuclei provided that   
neutrons can be excited to a unique-parity orbit, reducing a proton 
spin-orbit splitting crucial for the deformation or {\it vice 
versa}.  Some variations of this mechanism are also of great interest.
Type II shell-evolution mechanism should have been included in shell-model 
calculations in the past, for example \cite{lenzi,ni68_sorlin,ni68_dijon}, 
provided that tensor-force component was included properly and a 
sufficiently large model space was taken.  Nevertheless, its explicit recognition should 
help us to understand the underlying physics and foresee its impacts on structure 
issues in various regions of the nuclear chart including heavier and more exotic ones 
to be explored.    

The prolate band, being discussed, comes down to the 0$^+_2$ and 2$^+_2$ states 
as $N$ increases from 40 to 42 or 44 (see Fig.\ref{fig3}(g,i)).
The observed 2$^+_2$ level of $^{70}$Ni is as low as 2 MeV, 
which is reproduced well by the present calculation, as shown in Fig.~\ref{fig1}(b).     
The fact that this level has not been reproduced by calculations 
with limited configurations \cite{jun45,jj44b} suggests that the $Z$=28 core is broken.

Moving to $^{74}$Ni, we observe that Fig.~\ref{fig3}(j,k) exhibit another interesting
pattern.   The distribution of the circles becomes wide in both magnitude 
and $\gamma$ direction, {\it i.e.}, triaxiality.  A similar distribution is obtained also for
the 2$^+_2$ states, and the situation is the same for $^{76}$Ni.  This pattern 
resembles the critical-point-symmetry $E$(5) \cite{E5}.  
   
Finally, we come to $^{78}$Ni, a doubly-magic nucleus.  
Figure~\ref{fig3}(l) shows the PES and wave function distribution.
The PES has a spherical minimum which is very flat, as shown by the blue area of 
Fig.~\ref{fig3}(l). One sees that the circles almost fill the entire flat area.
This interesting pattern is seen also for the 2$^+_1$ state, reflecting a particular fluctuation.
This fluctuation is much narrower in $^{68}$Ni, where the E2 excitation from 
the ground state goes to very high 2$^+$ states, despite the low-lying 2$^+_1$ level.
On the other hand, the overlap probability with the closed shell is 60\%, 53\%, 75\%
for $^{56,68,78}$Ni, respectively, similar to \cite{sieja}.    Thus, $^{56,68,78}$Ni 
show very interesting variations regarding the appearance of the
magicity, which deserve further investigation.

In summary, the advanced MCSM calculations 
present intriguing variations of shapes, analyzed in terms of intrinsic shapes.  
The shell evolution inside the same nucleus, called Type II shell evolution for
clarification, 
can occur and can enhance shape coexistence.  
While Type I shell evolution appears in the $5/2^-$-$3/2^-$ inversion of Cu isotopes 
\cite{vmu,flanagan}, not only particles in $g_{9/2}$ but also holes in $f_{5/2}$  
contribute to the Type II case, making this case even more prominent.  
In addition, various shape evolutions are seen as $N$ changes, with notable fluctuations.  
Thus, the shapes of exotic nuclei provide us with many new features.
In stable nuclei, the shape has often been discussed as functions of $N$ and $Z$. 
For instance, shape evolution from vibrational to rotational nuclei as $N$ increases.   
Such a simple classification may no longer be appropriate in exotic nuclei.       
Type II shell evolution is expected in heavier nuclei, as unique-parity orbits 
come down, and also because of the robust tensor-force effect \cite{vmu,tsunoda}.

We thank Prof. B.R. Barrett for valuable comments.
This work was in part supported by MEXT Grant-in-Aid for 
Scientific Research~(A) 20244022.  
This work has been supported by HPCI (hp120284 and hp130024), 
and is a part of the RIKEN-CNS joint research project on large-scale nuclear-structure calculations.
Y.T. acknowledges JSPS for Research Fellow (No. 258994).  


\end{document}